\documentclass[smallextended,graphicx,amssymb,floatfix]{revtex4}

\begin{document}

\title{Weak value controversy}
\author{  Lev Vaidman}
\address{ Raymond and Beverly Sackler School of Physics and Astronomy,
 Tel-Aviv University, Tel-Aviv 69978, Israel}
\begin{abstract}
Recent controversy regarding the meaning and usefulness of weak values is reviewed. It is argued that in spite of recent statistical arguments by Ferrie and Combes, experiments with anomalous weak values provide a useful amplification techniques for precision measurements of small effects in many realistic situations. The statistical nature of weak vales was questioned. Although measuring weak value requires an ensemble, it is argued that the weak value, similarly to an eigenvalue, is a property of a single pre- and post-selected quantum system.
\end{abstract}
\maketitle


\section{Aharonov, Albert and Vaidman paper}

The concept of the weak value was introduced almost 30 years ago by Aharonov, Albert and Vaidman (AAV) \cite{AAV}. From the publication of the Letter ``How the Result of a Measurement of a Component of the Spin of a Spin-$\frac{1}{2}$ Particle Can Turn Out to be 100?''
  and until today it continues to be in the center of a hot controversy. Here I will review some of its controversial aspects and will clarify my point of view.

In \cite{AAV} the weak value was defined as the outcome of the usual measuring procedure with weakened coupling performed on pre- and post-selected ensembles of
quantum systems. The weakness condition was that the coupling  does not change significantly the quantum state of the system.
 The concept was defined in the framework of the two-state vector formalism which describes pre- and post-selected systems by both forward and backward evolving quantum states.
The measurement interaction has to be weak enough not to change significantly these  states, where ``significantly'' means that their scalar product remains approximately the same at all times during the measurement interaction.

 It was shown that in the standard  von Neumann model of measuring a variable $A$, in the limit of weak coupling, the quantum state of the pointer after the post-selection, $\Psi (q)$, is ``shifted''  by the weak value $A_w$
\begin{equation}\label{shift}
\Psi (q) \rightarrow \Psi (q-A_w).
\end{equation}
 The reason for apostrophes is that $A_w$ is not like $q$, apart from an unmentioned scaling factor, it might be a complex number, in which case (also not mentioned) renormalization is needed.
For a
pre- and post-selected system described at time $t$ by the two-state
vector  $ \langle \phi |~~|\psi\rangle $ \cite{TSVF}, the weak value
is defined as
\begin{equation}\label{wv}
A_w \equiv \frac{ \langle{\phi} \vert A \vert\psi\rangle}{
\langle{\phi}\vert{\psi}\rangle } .
\end{equation}
The real part of  $A_w$, which is the average reading of the pointer of the standard measuring device,  might be much larger than all eigenvalues.

Shortly after appearance of  the AAV paper,   Duck,    Stevenson, and   Sudarshan wrote \cite{Duck}: ``One's initial reaction is that this is impossible. This prejudice is reinforced when one finds that AAV's paper contains several errors.'' Leggett and Peres published (critical) Comments in PRL \cite{AAVLeg,AAVPer}. They could not accept that a value associated with a physical variable $A$ can be anything different from some eigenvalue of $A$. But continuation in \cite{Duck} was:``Nevertheless, after a careful study, we have concluded that AAV's main point does have validity'' and after appearing of our reply to Peres and Leggett \cite{AAVrep}, Duck,    Stevenson, and    Sudarshan added: ``A new manuscript by Aharonov and Vaidman \cite{AAVrep} clarifies the mathematical example originally presented in Ref. l. We refer the interested reader to this paper, and withdraw our earlier criticism of this example.''

In his Comment, Leggett wrote:``In a true measurement, by contrast, the measured value tells us much more than just the effect of the system on the measuring device.''
It shows that this controversy is about semantics. For me, the main relevant thing regarding a physical variable of a system, is how it affects other systems. And we stressed in the reply \cite{AAVrep} that
``{\it any measuring procedure of a   physical variable the coupling can be made weak enough such that the effective value of the variable for a preselected and postselected ensemble will
be its weak value}.''  Moreover, since the result does not rest on the specific form of the interaction, it needs not be a measurement interaction. The only requirement is the weakness of the interaction.

Any weak enough coupling to a variable $A$ is an effective coupling to  $A_w$. For a coupling to a continuous variable we obtain the ``shift'' (\ref{shift}), and more generally, for a weak coupling to any variable,  in the interaction Hamiltonian   the operator $A$ can be replaced by the c-number $A_w$ \cite{WMWV}.

\section{Amplificaiton}

 The hope for practical applications  of the weak measurement procedure was expressed in the conclusions of \cite{AAV} which pointed out that with small scalar product $\langle{\phi}\vert{\psi}\rangle $ we get ``tremendous amplification'' of small effects. The first experiment \cite{Ritchie} showed a factor  of 20. It was the work of Hosten and Kwiat \cite{Kwiat} twenty years later that used the amplification effect for observing the  spin Hall effect for light, which brought the AAV amplification scheme to the center of current research. This tiny effect  had not been observed before by any other means. Shortly after, weak value measurement techniques allowed Dixon {\it et al.}  \cite{Dixon} to measure an unimaginably small rotation of a mirror and many more implementations of the AAV method were reported \cite{Stru,Xu,Zhou,JaMe,Boyd,cycle,WVAnonlinear}.

This activity brought    a new controversy. Ferry and Combes (FC) posted a preprint titled ``Weak values considered harmful'' \cite{harm}. Robert Garisto, the Editor of Physical Review Letters chose this submission as particularly interesting, but softened its title: ``Weak Value Amplification is Suboptimal for Estimation and Detection'' \cite{FC1}.
 Numerous works  were published praising and criticising the AAV scheme as a method    for parameter estimation \cite{Kedem,Knee,JoHo,Das,Lee,PaBrPRL,ZDW,Nish,Al-Da,PaBrPRA,SuTa,ViHo,DenkmayrMatterWaveIF,Torr,Zhang,KneGaug,Liu,ParkSpe,PangBrun,Mirhos,Gross,Alves}. Indeed, theoretical analysis is complicated and depends on the models of noises in the experiment. However, I find that most of these sophisticated  analyses are unnecessary for  explaining why, in spite of the ``statistically rigorous arguments'' of  Ferrie and  Combes  \cite{FC1}, the AAV amplification scheme was  useful in numerous experiments \cite{FCcom1}.
 The explanation is that the assumptions in their statistical analysis are irrelevant for many realistic experimental situations.

I found the main erroneous  assumption which led Ferrie and Combes to their incorrect conclusions  thanks to my  direct involvement in weak measurement experiments \cite{Xu,Danan}. The limiting factor in these and other experiments is not the number of preselected quantum systems  (photons) considered by Ferrie and Combes, but the number of detected, post-selected photons. The saturation of the detectors generally happens much before the power limitation of the laser source kicks in. Thus, the low probability of the postseletction, the main negative factor in experiments with anomalously large weak values, is not relevant. In fact, I also have been involved in a weak value measurement experiment recycling photons which were not post-selected \cite{VKwi}, but the results only convinced  me that it is an unnecessary complication of the experiment.

The argument made in my comment \cite{FCcom1} was recently developed  by Harris, Boyd and Lundeen  in PRL: ``Weak Value Amplification Can Outperform Conventional Measurement in the Presence of Detector Saturation'' \cite{HBL}. Ironically, their Letter includes the statement ``saturation alone does not confer an advantage to the WVA'' which is technically valid due to their assumption of a specific saturation model and unrealistic ideal noiseless situation. The main point of their Letter was what I stated in my Comment: Ferrie and Combes' calculations, as well as a few other similar results, are  not relevant for real experiments.

\section{Classical analog to weak value}

Ferrie and Combes took the controversy about anomalous weak values even further,  publishing another PRL with  distinction \cite{FC2}. It has a provocative title: ``How the Result of a Single Coin Toss Can Turn Out to be 100 Heads''. In this  Letter,   FC claimed to show ``that weak values are
not inherently quantum, but rather a purely statistical feature of pre- and post-selection with disturbance.'' To prove their point, they  presented a purely classical situation with a coin toss which is supposed to be analogous to the example presented in the first publication of the weak value which has the title: ``How the Result of a Measurement of a Component of the Spin of a Spin-$\frac{1}{2}$ Particle Can Turn Out to be 100'' \cite{AAV}.

In my view, the analogy is  an illusion \cite{FCcom2}. The weak value of a variable of a system is defined by pre-selected and post-selected states of the system. The weak value of 100 for the spin $z$ component of a particle appeared for the particular pre- and post-selected spin states:
\begin{eqnarray}
 \nonumber  |\psi \rangle &=& \cos \frac{\alpha}{2}|\uparrow_x \rangle + \sin \frac{\alpha}{2}|\downarrow_x \rangle, ~~~~~\tan\frac{\alpha}{2}=100,\\
 |\phi \rangle &=& |\uparrow_x \rangle.
\end{eqnarray}
The number ``100'' appears due to the almost opposite directions of pre- and post-selected spins and specified by the parameter $\alpha$ of the pre-selected state. It does not depend on a particular disturbance of the measurement: every weak enough coupling to the spin will show $(\sigma_z)_w=100$. The disturbance of the measurement might distort the weak value, it does not specify it. The number 100 is obtained in the limit of vanishing disturbance.

 In contrast, in  the example of Ferrie and  Combes, the initial state is ``1'' and the final state \mbox{is ``-1'',} their classical system does not have enough complexity to define different numbers. There are only four possible pre- and post-selections, so we can get only four possible ``weak values'' of a given variable.
Ferrie and Combes got  the value   100   by playing with the definition of disturbance in their  ``weak'' measurement. They could equally well get value 1000 for the same pre- and post-selection. There is nothing in their construction analogous to (\ref{wv}) that  provides a functional dependence on the pre- and post-selected  states of the system. The continuum of classical ``analog of weak values'' is obtained by tailoring the interaction. The difference between the AAV and FC is not just quantum versus classical, the setups are conceptually different, so there cannot be an analogy between the two cases.

Apart from my Comment \cite{FCcom2} on the second PRL of Ferrie and Combes there were many more: \cite{CAhRo,CCoh,CHof,CBrod,CSoc,CRom,Mund}.
 PRL chose to publish only the Comment of Brodutch \cite{CBrod}. In my view, it was the most convenient choice for Ferrie and Combes. It pointed out that the classical model of the FC Letter had a technical mistake: a measurement of a variable that can have values $\pm 1$ could not yield 100. Probably, the best reply of FC would be: sure, Brodutch is right, the measurement procedure is not legitimate, but the error is exactly the same as in the  AAV quantum measurement procedure!

 FC in their reply \cite{CBrodR}, instead, made again the connection to the work of Garretson {\it et al.} \cite{Gar}  (based on \cite{Wis04}) discussing  ``weak-valued probability distribution of momentum transfer'' in which-way experiments.  The  term ``weak value of probability'' has rigorous definition as the  weak value of  the projection operator. In this experiment a weak value of a projection operator on a particular momentum was measured. The important difference in this work relative to the weak value of a standard weak measurement is the presence of additional which-way measurement, not related to the weak measurement of momentum. It is this additional  which-way measurement which caused the disturbance. The disturbance does not go to zero with the limit of vanishing coupling of the weak measurement. Apparently, the analogy to this disturbance   lead to ``100 heads'' in the FC example. The number 100  did not  come  from pre- and post-selection of the coin. See more analysis of disturbance in weak measurements in \cite{Ips,Dressel}.

I doubt that a correct classical analogy of the AAV experiment exists, but it at least can be formulated when we consider a toss of a real classical coin with pre- and post-selected states specified by their actual orientation in space, compare with recent proposal \cite{Gond}. The space of pre- and post-selected states then is rich enough for a functional relation (\ref{wv}).

 \section{Weak value as a contextual value}

  Another line of argument against FC's claim was the result  by Pusey \cite{Pus} who showed that   anomalous weak values constitute proofs of the incompatibility of quantum theory with noncontextual ontological models \cite{Spek}. This result has recently been demonstrated experimentally \cite{Pian}.
 The connection between weak values and contextuality was pointed out by Dressel {\it et al.} \cite{DresCV1,DresCV2} who introduced ``contextual values''  and viewed the weak value as an example of a contextual value.  ``The idea behind contextual value stems from the observation that the intrinsically measurable quantities in the quantum theory are the outcome probabilities for a particular measurement setup.''

 Although the conclusion of this approach is what I strongly believe: anomalous weak values cannot be explained in the framework of classical statistical theory, I am very far from accepting the connection between contextual values and weak values. At the price of accepting parallel worlds, I can view quantum theory  as a deterministic theory \cite{QMD}. And I disagree that  it is  based on outcome probabilities:
  apart for predicting (the illusion of) probabilities for outcomes of experiments, quantum mechanics  makes   numerous definite predictions: spectrum of atoms, etc.

 If we insist on considering probabilistic theories, then there is a way to introduce weak values in a classical theory, but a very specific one, a classical theory with an epistemic restriction \cite{Karan}. In this theory, as the authors point out, ``anomalous weak values do not appear in our analysis, as all observables in our model possess an unbounded spectrum. Consistent with the results of \cite{Pus}, our model is also noncontextual: the ERL [epistemically restricted Liouville] mechanics provides an explicit noncontextual ontological model for all procedures described here.''

Contextual value techniques lead to a natural definition of a general conditioned average that converges uniquely to the quantum weak value in the minimal disturbance limit.
And I find it of interest that it helps to answer the criticism of Parrott \cite{Parr,DresCV3}. However, I have another argument which allows one to avoid dealing with statistics. In the next section I will argue that the weak value can be considered beyond its statistical meaning. And if the nature of weak values is not statistical, then statistical analyses are not relevant.

 \section{Weak value as a property of a single system}

Recently, together with Harald Weinfurter and his group in Munich, we brought another theoretical argument demonstrated in an actual experiment which  refutes any attempt to find a classical statistical analog of the weak value \cite{EWEV}. The argument is that the weak value is a property of a single pre- and post-selected quantum system.  Thus it cannot have an analogy as a statistical property of an ensemble.

Weak values were introduced as outcomes of weak measurements \cite{AAV}, which have large uncertainty in the pointer position.
Thus, in experiments, the weak value is obtained as a statistical average of the pointer readings.
Even among proponents of this concept, the weak value is frequently understood as a  mere generalization of the expectation value for the case when the quantum system is post-selected, i.e., a conditional expectation value \cite{wiseman,review}.

Contrary to the classical case, if we are given a single system in a known pure quantum state, we cannot test this fact with certainty. Still there is some certainty about this situation, we know that a projection measurement on this state will succeed with certainty. This what makes this situation not statistical. If we have a system with a known eigenvalue of a variable, we know with certainty the result of a measurement of this variable. This is not the case if we know that the system is described by a particular known expectation value of a variable. What we have found in \cite{EWEV}, is that when the system is described by a weak value, it interacts with other systems almost identically to the system described by numerically equal eigenvalue, but significantly different from the case of numerically equal expectation value.

What we want to compare is the operational meaning of a weak value, an eigenvalue and an expectation value.
While the definition of the weak value is based on the pure two-state vector with states $ |\psi \rangle$ and $ |\varphi \rangle$  considered at a particular time $t$, its operational meaning relies on interactions with other systems which create entanglement.
The way to deal with this problem is  to consider a short period of time around time $t$ to evaluate the action of the system on other systems. During this small time the entanglement can be considered negligible, but then the effect is very small too. To observe it we need an ensemble.  We attribute properties to each system individually, but for testing our claim we will use an ensemble.

We consider a standard measuring procedure described by interaction  Hamiltonian ${H_{\rm int}= gA P}$.
We assume that at time $t=0$, the system was prepared in state $|\psi \rangle$ and shortly after, at time $t=\epsilon$, was found in state $|\varphi \rangle$.
The pointer at time $t=0$ is in a Gaussian state $\Phi_0$.
For a comparison of different cases, we consider the pointer state at time $t=\epsilon$, after the interaction with   the integer spin observable
$
A\equiv \sum_j j | j \rangle \langle j |$ .
If the spin state is the eigenstate $|1\rangle$, i.e. the variable has the eigenvalue $A=1$, then at time $t=\epsilon$, independently of the result of the post-selection measurement, the pointer state is shifted:
\begin{equation}\label{PhiF}
\Phi_{\rm e}=\mathcal{N} e^{-\frac{\left(Q-g\epsilon \right)^{2}}{4\Delta^{2}}}.
\end{equation}

To compare the various cases we evaluate the effect of the interaction by calculating the distance between quantum states  expressed by the Bures angle.
The distance between the initial state of the measuring device $\Phi_{\rm 0}$ and the final state (\ref{PhiF}) is
\begin{equation} \label{Bures_weak}
D_A\left(\Phi_0,\Phi_{\rm e}\right)\equiv\arccos \left|\left\langle \Phi_0|\Phi_{\rm e}\right\rangle \right|=\frac{g\epsilon}{2\Delta}+\mathcal{O}\left(\epsilon^3\right).
\end{equation}

Consider now a pre- and post-selected system with $A_w=1$, but in which both pre-selection and post-selection do not include the eigenstate $|1\rangle$.
A two-state vector which provides this weak value is
\begin{equation}\label{TSV}
\langle\varphi|~~|\psi\rangle=\frac{1}{\sqrt 5}\left(\langle-1| - 2 \langle0|\right)~~~\frac{1}{\sqrt 2}\left(|-1\rangle+ |0\rangle \right).
\end{equation}
After the post-selection, the state of the pointer variable is
\begin{equation}
\Phi_{\rm w}=\mathcal{N}(\epsilon)(2e^{-\frac{Q^{2}}{4\Delta^{2}}}-e^{-\frac{(Q+g\epsilon)^{2}}{4\Delta^{2}}}) \approx \mathcal{N}^\prime(\epsilon) e^{\frac{ -Q^2g^2\epsilon^2}{4 \Delta^4}} \Phi_{\rm e}.\label{wvpointer}
\end{equation}
 $\Phi_{\rm w}$ is effectively a Gaussian centered around $A_w=1$ and is thus very close to $\Phi_{\rm e}$ as seen from the Bures angle
\begin{equation} \label{Phi2-Phi1}
D_A\left(\Phi_{\rm e},\Phi_{\rm w}\right)=\frac{g^2\epsilon^2}{2\sqrt{2}\Delta^2}+\mathcal{O}\left(\epsilon^4\right).
\end{equation}

The characteristic distance between states after the interaction for the time $\epsilon$ is approximately $\frac{g\epsilon}{2\Delta}$, so when the additional distance is proportional to $\epsilon^2$, it can be neglected.
Thus, in the limit of short interaction times, the pre- and post-selected system with some weak value interacts with other systems in the same manner as a system pre-selected in an eigenstate with a numerically equal eigenvalue.
Not only the expectation values of the positions of the pointers are essentially the same, but the full quantum states of the pointers are almost identical.

The situation changes considerably when the system is only pre-selected in a state with the expectation value $\langle A\rangle=1$, which, however, is not the eigenstate $|1\rangle$.
To show this, assume that the particle is in the state
\begin{equation}\label{psi3}
|\psi\rangle= \frac{1}{\sqrt 2}\left (| 0\rangle +| 2\rangle \right ).
\end{equation}
At time $t=\epsilon$, now without post-selection, the pointer system is not described by a pure state, but by a mixture.
The density matrix describing this mixture is
\begin{equation}
\rho_{\rm ex} \!=\!\frac{1}{2\sqrt{2\pi}\Delta}\left(e^{-\frac{Q^2+Q'^2}{4\Delta^2}}\!+\!e^{-\frac{(Q-2g\epsilon)^2 +(Q'-2g\epsilon)^2}{4\Delta^2}}\right)\!.\label{mixed}
\end{equation}

The distance between $\rho_{\rm ex}$ and $\Phi_{\rm e}$, the state of the pointer after coupling to an eigenvalue, is
\begin{equation} \label{overlap_mixture}
D_A\left(\Phi_{\rm e},\rho_{\rm ex}\right) \equiv \arccos(\sqrt{\langle \Phi_{\rm e} |\rho_{\rm ex} | \Phi_{\rm e} \rangle})=\frac{g\epsilon}{2\Delta}+\mathcal{O}\left(\epsilon^3 \right).
 \end{equation}
This is a significantly larger distance than (\ref{Phi2-Phi1}).
In fact, the distance (\ref{overlap_mixture}) is of the same order as (\ref{Bures_weak}) and cannot be neglected for small $\epsilon$.

While the pointer states (\ref{wvpointer}) and (\ref{mixed}) for a small enough $\epsilon$ correspond to similar probability distributions, they are fundamentally different.
As in the case of an eigenstate (\ref{PhiF}), the final pointer state (\ref{wvpointer}) corresponds to a shift of the original distribution given by a single number, the weak value.
When the system is prepared in a superposition of eigenstates,
 the result is  a  mixture of  two independent pointer distributions centered around the values $0$ and $2$, which cannot be described by a single parameter anymore.

Our demonstration of  the weak value as a property of a single pre- and post-selected system shows that recent classical statistical analogies of weak values \cite{FC2} which can be formulated only given an ensemble, are artificial.

\section{Conclusion}

Quantum theory is about century old, but its foundations are still under a hot debate.  The quantum phenomena are very different from the classical picture, so there was a tendency in the early days to accept that quantum reality is not understandable in principle, and some questions should not be asked. One such question is the description of a pre- and post-selected quantum system. The weak value is a property of such a system and the standard formalism, without a backward evolving quantum state, lacks this concept.
It is not that we cannot discuss the interaction of a pre- and post-selected system in the standard formalism,  it is just much more difficult and lacks a transparent picture, since it has to involve entanglement with other the systems.

As one can see from the unusually long reference list which mostly consists of papers claiming contradictory statements, we are far from reaching a consensus about the meaning of weak values, and there are several other related controversies. What is the status of counterfactual statements about pre- and post-selected quantum systems \cite{RFTS,DefTS}? What can be said about the past of quantum particles \cite{past}? (I do not present the list of relevant references here.) Can weak values be measured strongly \cite{SWV,ComSWV}?

In this paper I just covered the controversies raised by the two Physical Review Letters of Ferrie and Combes \cite{FC1,FC2}.  Although, as a referee, I think that these Letters should have never been published (not to mention receive distinctions), in retrospect I admit that the controversy they raised brought deeper understanding of quantum mechanics and that this controversy can be considered as a part of a second quantum revolution. Our discovery that the weak value is more like an eigenvalue than an expectation value and that it is a property of a single system and not of an ensemble is far from being accepted, so the revolution is still in progress.

This work has been supported in part by the Israel Science Foundation Grant No. 1311/14,
the German-Israeli Foundation for Scientific Research and Development Grant No. I-1275-303.14.

\end{document}